\newcommand{\go}   {\tilde{g}}

\newcommand{\sbx}[1]{\tilde{b}_{{#1}}}
\newcommand{\sq}[1]{\tilde{q}_{{#1}}}
\newcommand{\se}[1]{\tilde{\ell}_{{#1}}}

\newcommand{\nn}[1]{\tilde{\chi}^0_{{#1}}}

\newcommand{\pb}{~{\ensuremath\rm pb}}
\newcommand{\ifb}{~{\ensuremath\rm fb^{-1}}}
 
\documentclass[
    ,final            
  ]
  {aipproc}

\layoutstyle{6x9}

\begin{document}

\title{Spins in Gluino Decays} 


\author{Alexandre Alves}{
        address={Instituto de F\'\i sica, Universidade de S\~{a}o Paulo,
                 S\~{a}o Paulo, Brazil}}
\author{Oscar \'Eboli}{
        address={Instituto de F\'\i sica, Universidade de S\~{a}o Paulo,
                 S\~{a}o Paulo, Brazil}}
\author{\underline{Tilman Plehn}}{
        address={Heisenberg Fellow, MPI for Physics, Munich, Germany \\
                 and SUPA, School of Physics, University of Edinburgh,
                 Scotland}}

\begin{abstract}
  To unambiguously claim that we see a gluino at the LHC we need to
  prove its fermionic nature. Looking only at angular correlations we
  can distinguish a universal extra dimensional interpretation of a
  gluino cascade decay (assuming a bosonic heavy gluon) from
  supersymmetry. In addition to the known lepton--hadron asymmetries we
  also use of purely hadronic correlations in the gluino--sbottom
  decay chain.
\end{abstract}

\maketitle


Looking for charged leptons in gluino decays we can test
supersymmetric QCD: like--sign dileptons occur in gluino pair
production because the Majorana gluino can decay to $q \sq{}^*$ or
$\bar{q} \sq{}$, where the (anti)squark decay yields a
definite--charge lepton~\cite{likesign}.  This recipe for identifying
a strongly interacting Majorana fermion (called gluino) has a loop
hole. If the particle responsible for a gluino--like cascade decay is
a boson~\cite{ued, early} with adjoint color charge, like--sign
dileptons naturally occur as well. We need to show that the gluino
candidate is a fermion~\cite{us}. Such a determination of quantum
numbers of new particles is a necessary addition to measuring their
Lagrangian's parameters at LHC~\cite{edges, sfitter, gordi}, because
these studies implicitly assume that we know the spin of all new
particles.

Depending on the mass spectrum, the gluino mass can be determined in
the cascade decay $\go \to b \sbx{1}^*/\bar{b} \sbx{1}$, where the
light sbottom decays through $\sbx{1} \to \nn{2} \to \se{} \to
\nn{1}$~\cite{edges,mgl}.  This well-studied decay we use for our spin
analysis.

In the similar case of a squark decay we know how to show that the
$\sq{L}$ is indeed a scalar~\cite{barr,smillie}: first, we assume (for
gluino decays to bottoms we know) that all SM particles radiated from
the cascade decay are fermions.  To determine the spin of the decaying
particle all we have to do is compare the SUSY cascade with another
interpretation, where the intermediate states have the same spin as
their SM partners. Such a model are Universal Extra Dimensions
(UED)~\cite{ued,early}. There are many ways to discriminate `typical'
UED and SUSY models. Direct spin information is generally extracted
from angular correlations or Lorentz--invariant normalized invariant
masses. For example, if we knew which of the leptons in the squark
decay is radiated right after the quark we could use $m_{q\ell}$ to
distinguish UED from SUSY cascades~\cite{edges,barr,smillie}. Instead,
the $\sq{L}$ cascade analysis relies on an asymmetry of squark vs.
anti--squark production with a gluino~\cite{smillie}.  For our gluino
decay tagged bottoms can include a lepton, which allows us to
distinguish $b$ and $\bar{b}$ on an event--by--event basis.

\subsubsection{Lepton--Bottom Correlations}
\label{sec:sps}

For a quantitative study we choose the parameter point
SPS1a~\cite{sps}. The NLO production rates are $7.96 \pb$ for $\go\go$
and $26.6 \pb$ for $\sq{}\go$~\cite{prospino}. To avoid combinatorial
backgrounds we require one gluino decay through the short cascade with
a light--flavor squark decaying into one light--flavor jets and the
LSP. For the second gluino we require two tagged bottom jets in the
long cascade through a slepton. This selection means that it is
straightforward to also include the large associated
$\sq{}\go$ production, where the squark decays to a jet and the LSP.
The dominant SM background is obviously $t\bar{t}$+jets.  All
backgrounds include uncorrelated leptons from independent decays.  An
efficient way to eliminate these backgrounds beyond the level $S/B
\sim 1$ is to subtract the measured opposite flavor dileptons from the
same flavor dileptons~\cite{Gjelsten:2004ki}.  Because the precise
prediction of the remaining tiny backgrounds is beyond the scope of
this paper we will not include them in our analysis.

\begin{figure}[t]
  \includegraphics[width=13.0cm]{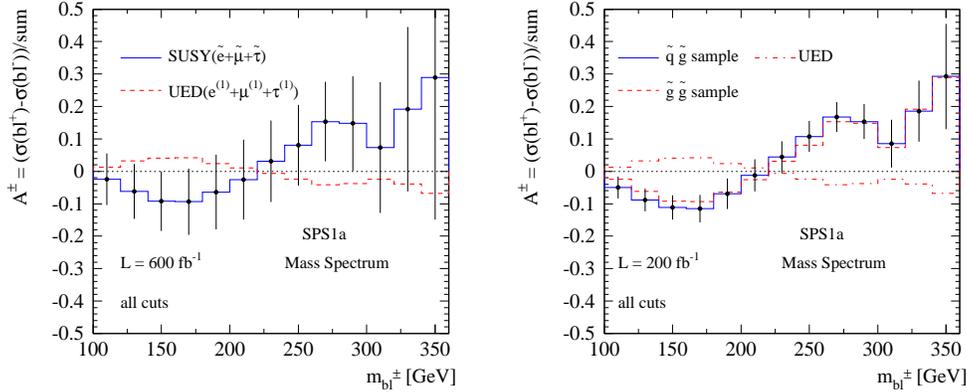}
  \caption{$A^\pm(m_{b\ell})$ for the gluino-pair channel after 
    acceptance cuts and for the gluino--pair and squark--gluino
    channels after background rejection cuts.}
\label{fig:complete}
\end{figure}

For SPS1a parameters, the mass hierarchy has a favorable impact on the
decay jet momenta. Just picking the harder of the two bottom jets we
could distinguish between near and far jets. We choose to ignore this
spectrum dependent approach in favor of the general method of
distinguishing bottom and anti--bottom jets.  For the simulation we use
UED in Madevent/Smadgraph~\cite{smadgraph}. Detector effects and
tagging efficiencies yield an additional $0.11$ dilution factor for
the signal~\cite{btags}.  We assume the SPS1a spectrum also for the
UED particles and normalize the rates to the SUSY case. All
information left in the $m_{b\ell}$ shape are now angular
correlations. We construct the asymmetry
\begin{equation}
  A^\pm(m_{b\ell}) = 
  \frac{d\sigma/dm_{b\ell^+} - d\sigma/dm_{b\ell^-}}
  {d\sigma/dm_{b\ell^+} + d\sigma/dm_{b\ell^-}}  
\label{eq:asymm}
\end{equation}
In Fig.~\ref{fig:complete} we show some results after acceptance and
background rejection cuts. The asymmetry is not biased by the harder
cuts. The $\sq{}\go$ and $\go\go$ contributions can indeed be added
naively.

The details of the gluino decay chain reveal something else~\cite{us}:
two leptons in the cascade usually come from an intermediate
light--flavor slepton.  Alternatively, the decay can proceed through
staus. For SPS1a the lighter selectron/smuon is mostly right handed
whereas the lighter stau is mostly left handed, due to the
renormalization group running and the fairly large $\tan\beta = 10$.
This means the stau contribution to the mass asymmetry is opposite to
the selectron/smuon contribution and that it can wash it out. Luckily,
its numerical impact can be suppressed because leptons from tau decays
are significantly softer.

\subsubsection{Bottom--Bottom Correlations} 
\label{sec:had}

\begin{figure}[t]
  \includegraphics[width=.48\textwidth]{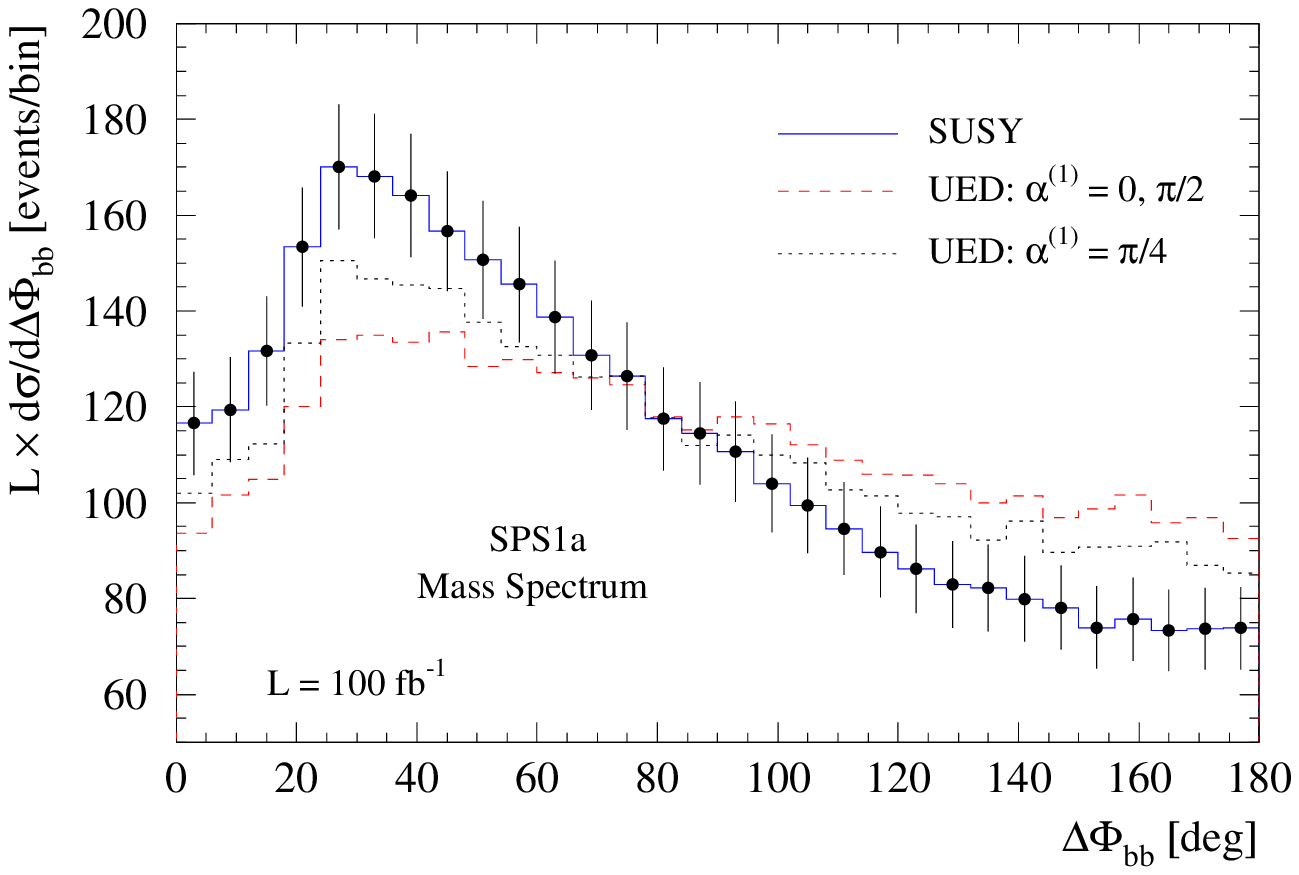} \hspace*{5mm}
  \includegraphics[width=.48\textwidth]{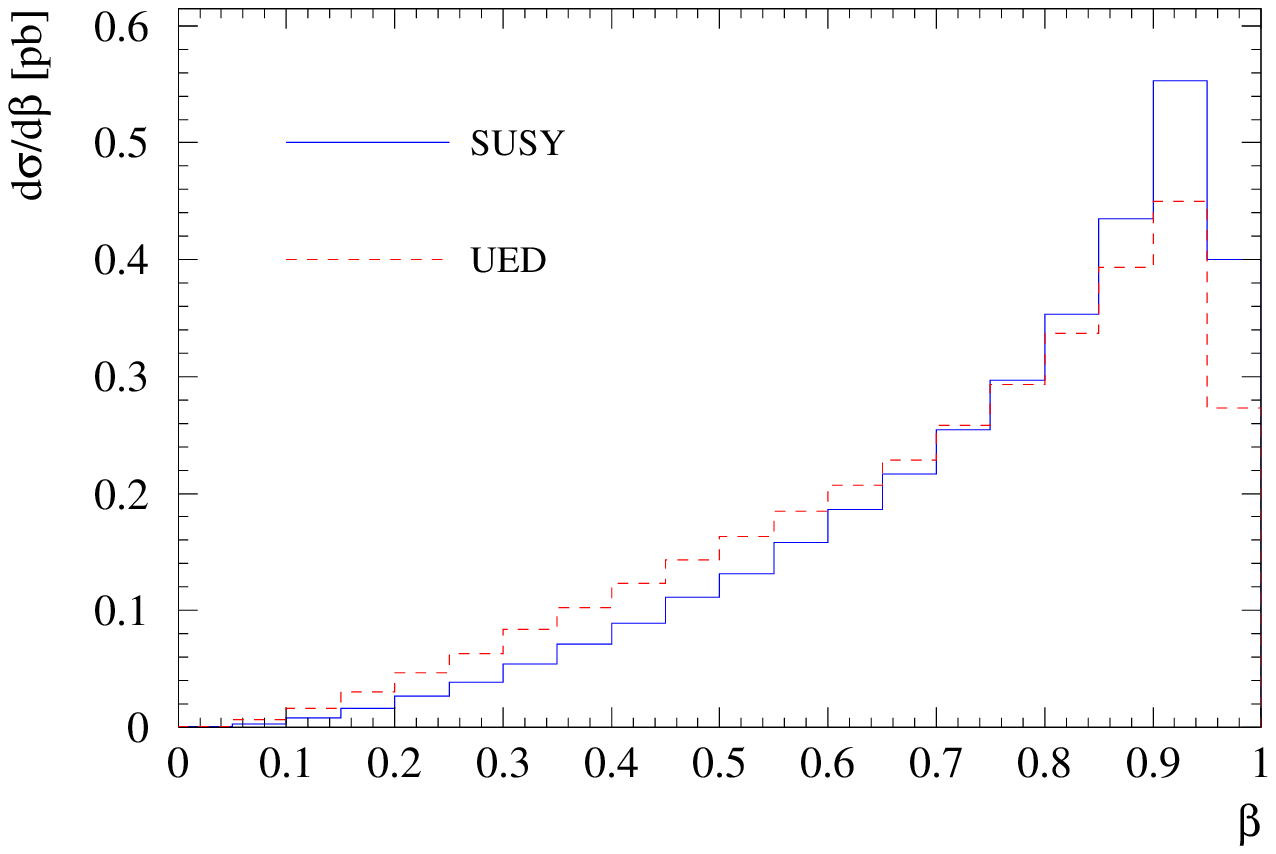} 
  \caption{Left: Azimuthal angle between the two bottom jets. 
    Right: boost of the gluino and heavy KK gluon.}
\label{fig:phibb}
\end{figure}

The correlation between a lepton and a bottom jet is only one of the
distributions we can use.  Unfortunately, just as for squark decays
purely leptonic distributions are not as useful as mixed lepton--jet
correlations~\cite{smillie}. For the gluino decay chain we can build
purely hadronic correlations.  They have the advantage of being
independent of the $\nn{2}$ decay, which can be complicated by
intermediate gauge bosons or three-body decay kinematics.

In Fig.~\ref{fig:phibb} (left) we present the distribution
$d\sigma/d\Delta\phi_{bb}$, which exhibits a distinct behavior for
SUSY and UED decay chains and can be used to construct an asymmetry:
\begin{equation}
 \frac{\sigma(\Delta \Phi_{bb}< 90^o)-\sigma(\Delta \Phi_{bb}> 90^o)}
      {\sigma(\Delta \Phi_{bb}< 90^o)+\sigma(\Delta \Phi_{bb}> 90^o)}
\end{equation} 
It assumes small values $0.08\pm 0.02$ for the UED spin assignment.
For the SUSY interpretation it is significantly larger $0.24\pm 0.02$.
The quoted errors are statistical errors for the combination of the
gluino--pair and gluino--squark production channels with an integrated
luminosity of $100\ifb$. Using the $\sbx{2}$ contribution to the decay
chain~\cite{us} we can see that the different behavior shown in
Fig.~\ref{fig:phibb} is mostly due to the boost of the heavy gluino or
KK gluon. This difference we show in Fig.~\ref{fig:phibb} (right).

\subsubsection{Outlook} 

Proving the presence of a Majorana gluino is the prime task for the
LHC to show that new TeV--scale physics is supersymmetric. It has been
known for a long time that like--sign dileptons are a clear sign for
its Majorana nature~\cite{likesign}. The loop hole in this argument is
to show that the gluino candidate is a fermion.

Using a set of lepton--bottom~\cite{smillie} and bottom--bottom
asymmetries we distinguish between the SUSY and the UED cascade
interpretations and thus determine the spin of the gluino~\cite{us}.
However, we also find that the spin information in the decay
kinematics is always entangled with the left and right handed sfermion
couplings~\cite{sleptons,us}. \bigskip

{\sl Acknowledgments:} We would like to thank Martin Schmaltz, Gustavo
Burdman, Chris Lester, and Matthew Reece for their insightful
comments.
 

\end{document}